# Biomimetic Liquid Metal Cell

Jingyi Li,[1,2] Mengwen Qiao,[1,2] Minghui Guo,[1] Zerong Xing,[1,2] Yunlong Bai,[1,2] Ju Wang,[1,2] Yujia Song,[1,2] Ren Xu,[2] Xi Zhao,[3,4,*] Jing Liu[1,2,5,**]

[1] *State Key Laboratory of Cryogenic Science and Technology, Technical Institute of Physics and Chemistry, Chinese Academy of Sciences, Beijing 100190, China.*

[2] *School of Future Technology, University of Chinese Academy of Sciences, Beijing 100049, China.*

[3] *Tianjin Key Laboratory for Advanced Mechatronic System Design and Intelligent Control, School of Mechanical Engineering, Tianjin University of Technology, Tianjin 300384, China.*

[4] *National Demonstration Center for Experimental Mechanical and Electrical Engineering Education, Tianjin University of Technology, Tianjin 300384, China.*

[5] *School of Biomedical Engineering, Tsinghua University, Beijing 100084, China.*

\* Email: zhaoxi@mail.ipc.ac.cn, jliu@mail.ipc.ac.cn

## Abstract

Gallium-based liquid metals, as a broad category of emerging functional materials with unique physical, chemical, and biological properties, offer numerous possibilities for advancing intelligent systems. However, a basic query persistently remains for the complex liquid metal system: Is there a minimal functional unit that can fully capture its diversity of morphology and function? Cells, as the most basic structural and functional units of life, are small in scale but have complex structures, functions, and life activities. Analogous to nature, this article proposes the concept of liquid metal cells, and systematically explores their construction routes, sensing capabilities, motion behaviors, and potential applications. We first construct a multi-phase composite structure with liquid metal as the nucleus, ionic solution as the cytoplasm, and polymer as the cell membrane by developing a layered cryogenic molding method. Furthermore, we reveal that liquid metal cells exhibit inherently versatile responsive characteristics and self-adaptive behaviors to thermal, pressure, chemical, electrical, and magnetic fields, indicating "small world, vast potential". Based on these fundamental findings, we finally demonstrate the feasibility of utilizing liquid metal cells as sensors, fluidic valves, and material transport carriers in flow channels through dynamic control.

## 1. Introduction

Gallium-based room temperature liquid metals have emerged as revolutionary functional materials that attracted considerable attention due to their unique physicochemical properties and widespread applications in heat transfer,[1] flexible device,[2,3] reconfigurable systems,[4,5], catalytic synthesis,[6] environmental governance,[7] and biomedical technologies.[8] In addition, they exhibit excellent biocompatibility and environmental compatibility.[9,10] In recent years, the discoveries of fundamental effects on liquid metals have revealed their remarkable bio-autonomous and external-field responsive properties, such as self-driven soft robots,[11] transition-state machines,[12] energy absorption and conversion,[13] phagocytosis,[14] memory function,[15] and liquid-metal amoebae.[16] These behaviors closely resemble those of natural mollusks, such as cell migration, cellular deformation, metabolism, cell differentiation, and active transport.[17] The emergence of these biomimetic clues indicates that the liquid metal living matter system is expected to become one of the most promising answers for biology analogies, bridging the gap between life and non-life.[18,19] Moreover, it addresses the inherent limitations of conventional non-metallic artificial biomimetic systems, including those based on proteins,[20] polysaccharides,[21] hydrogels,[22] ionic conductors,[23] liquid crystal elastomers,[24] and silicone elastomers,[25] in replicating natural life processes. These limitations stem from the fundamental material properties, particularly in energy conversion, multi-field responsiveness, autonomous movement, environmental adaptability, and the scalability of functions and applications.[26-30] While these pioneering systems have provided valuable insights into artificial life, their material constraints often result in compromised performance in one or more of these critical aspects.

However, almost all the previously observed bio-like phenomena in liquid metals occur in centimeter-scale droplets due to the size limitations of surface tension-dominated fluid manipulation. The autonomous three-dimensional motion and deformation of large-scale liquid metal masses remain technically challenging. This scale limitation presents a significant constraint when trying to build a new living matter system compared to natural, which spans from molecular to interacting life. To overcome this constraint and expand the scale of the liquid metal living matter system, a bottom-up approach is essential.[31] This approach requires developing fundamental structural units that both exploit the unique properties of small-scale liquid metals and enable the assembly of larger, more complex functional systems through ordered organization.

Learning from nature, cells are the fundamental structural, functional, and biological units of life. Creating artificial cells with specific sensing capabilities through laboratory means would represent a significant breakthrough in the field of artificial life. The natural cellular structure, mainly consisting of a membrane, cytoplasm, and nucleus (**Fig. 1A**), provides valuable insights into the liquid metal living matter system. The realization of liquid metal living matter at various scales can be approached through the construction of minimal liquid metal cellular units. In this configuration, liquid metal droplets serve as the nucleus, providing basic perception and sensing functions. The surrounding aqueous solution acts as cytoplasm, facilitating interactions between the metallic core and the external environment. The elastic cell membrane isolates individual liquid metal units while maintaining the dynamical dominance of the electric double layer over the small-scale liquid metal nucleus, simultaneously allowing material exchange with the environment. It is necessary to emphasize here that this biosimilar cell structure and component composition also coincide with the aforementioned principle of constructing liquid metal micro-units rather than deliberately mimicking the structure of the cell just for the sake of biomimicry.

Based on these correlations between liquid metals and living cells, this study proposes the concept of biomimetic liquid metal cells (LMCs). As the fundamental structural and functional units of liquid metal living matter, it is an extensible and multifunctional platform that fully exploits the comprehensive properties of liquid metals, possessing the potential for bottom-up assembly into more complex functional entities and even systems. First, we elaborate on the structural design, material composition, and fabrication protocol of LMCs. Then, we explore the response of LMCs under various physical and chemical fields and disclose that LMCs have natural sensing functions and respond immediately to environmental changes. In addition, LMCs can also actively exhibit electrophilic and magnetotropic behaviors followed by directional movements. Finally, as an illustrative example, we demonstrate the potential application of LMCs as sensors, fluidic valves, and material transport carriers in a converging channel. The LMCs developed here significantly expand the current understanding of liquid metals while offering new possibilities for constructing more sophisticated liquid metal living matters, paving a promising way in making the artificial life system from bottom up.

## 2. Materials and Methods

### 2.1 Materials

Unless otherwise stated, all liquid metal mentioned in this study refers to pure EGaIn. EGaIn is prepared by mixing 75.5 wt% gallium and 24.5 wt% indium, and melting the mixture in a vacuum oven at 150°C for 6 hours.

Sodium alginate (90%) and calcium lactate hydrate ($C_6H_{10}CaO_6 \cdot xH_2O$, 98wt%) were purchased from Shanghai Macklin Biochemical Technology CO., Ltd.

Poly (3,4-ethylenedioxythiophene)-poly (styrene sulfonate) (PEDOT: PSS, 1.1% concentration in $H_2O$, without surfactants, and exhibiting high conductivity) was purchased from Shanghai Aladdin Biochemical Technology CO., Ltd.

Iron nanoparticles (99.9% metals basis, 100 nm average particle size) were purchased from Shanghai Macklin Biochemical Technology Co., Ltd.

Deionized water used for solution preparation was obtained from a ultrapure water system (Direct-Q 3 UV, Merck Millipore, Germany).

### 2.2 Preparation of SA-PEDOT: PSS composite gel solution

Dilute the PEDOT: PSS solution, and then add Sodium alginate (SA) powder to the mixture. After manual stirring until homogeneous, degas the solution under a vacuum. The final concentrations are 0.033 wt% PEDOT: PSS and 1.5 wt% SA.

### 2.3 Details and procedural steps of the layered cryogenic molding method

**Step 1**: Fill the target Ecoflex mold halfway with a 0.5 mol/L NaOH solution, lightly shake to remove bubbles, and freeze at -8 ℃ for 6 hours until the solution is completely solidified. Then, place a pre-frozen EgaIn (Eutectic Gallium and Indium, 75.5% gallium and 24.5% indium by weight percent) sphere in the center of the upper surface. This prevents heat release during the solidification of the liquid metal, which could otherwise melt the already-formed sodium hydroxide hemispherical ice block. After 1 hour of freezing at -8 °C, inject NaOH solution (0 to 1°C) into the mold until it is filled. These steps ensure that the liquid metal remains protected by a solution layer during subsequent membrane formation, preventing direct metal-membrane adhesion that would otherwise compromise the original structure and properties.

**Step 2**: After freezing the solution-liquid metal double-layer ice balls for 6 hours, remove them from the mold and place them in a 0°C SA-PEDOT: PSS composite gel solution (1.5 wt% SA, 0.033 wt% PEDOT:PSS).

**Step 3**: After ensuring the hydrogel solution evenly coats the ice balls through multi-directional rolling, quickly transfer them into a 4 °C 5 wt% calcium lactate (CaL) solution and keep stationary until the internal ice balls melt (about 5 minutes). This ensures complete crosslinking. However, it is crucial to avoid leaving the solution in contact with the gel for too long, as the differing osmotic pressures between the inside and outside could cause a reaction between NaOH and CaL, affecting the permeability and transparency of the cell membrane.

**Step 4**: Retrieve the formed LMCs from the CaL solution and rinse them quickly 3 times with deionized water to remove any residual CaL solution from the surface. Then, place the cells into a 0.5 mol/L NaOH solution to conduct follow-up experiments.

## 2.4 Preparation of pre-frozen liquid metal core

Target volumes of liquid metal droplets were dropped onto pre-cooled smooth ice blocks prepared by freezing 0.5 mol/L NaOH solution, causing phase transition and transformation into relatively smooth solid spheres.

## 2.5 Characterization of hydrogel layer cross-sectional morphology

The composite hydrogel membrane samples were prepared for SEM imaging using a combination of cryogenic fracturing and freeze-drying techniques to preserve their native microstructure. First, the samples were rapidly immersed in liquid nitrogen for 3 min to achieve cryogenic freezing, followed by mechanical fracturing to create a clean cross-sectional surface. Subsequently, the fractured samples were subjected to freeze-drying in a Benchtop Freeze Dryer (Advantage 2.0, SP VirTis, America) under vacuum conditions at −50°C for 12 hours to remove residual water while maintaining the structural integrity of the hydrogel network. The dried samples were then sputter-coated with a thin layer of gold (SPUTTER COATER 108auto, CRESSINGTON, UK) to enhance conductivity and reduce charging effects during SEM imaging. The hydrogel layer cross-sectional morphology images were captured using field emission environmental scanning electron microscopes (FESEM, QUANTA FEG 250, America) operated at an acceleration voltage of 15 kV, a working distance of 10 mm, and under high vacuum conditions.

## 2.6 Electrical testing of LMCs

All resistance values were obtained using the Kelvin four-terminal sensing method on a data acquisition system (DAQ970A, Keysight, America) unless otherwise noted.

The potential difference between the two electrodes during the motion was supplied by a DC power supply (HLR-3660D, Henghui, China).

**2.7 Temperature testing of LMCs**

The mean temperature of the LMC was determined by averaging the temperature values measured by three T-type thermocouples (KPS-T-T-30-SLE-1000-SMPW-G, Kaipusen, China) placed at the bottom, middle, and top of the LMC. All temperature values were obtained on a data acquisition system (DAQ970A, Keysight, America).

**2.8 Mechanical testing of LMCs**

Mechanical compression tests of LMC were performed on a universal stretching compressor (Model F105, MARK-10, America). The dimensions of two electrode plates constraining the longitudinal space of LMCs are 25.4 mm×76.2 mm×10 mm, with an effective conductive area of 10 mm×76.2 mm.

**2.9 Preparation of liquid metal ferrofluids**

Preweighed 0.5 mL EGaIn and 0.31 g Fe particles (particle diameter = 100 nm) and mixed them in a beaker filled with 6 mL of 2 mol/L HCl solution. Stirring the mixture gently with a glass rod for 15 minutes will allow the Fe particles to be completely internalized into the liquid metal. The desired magnetized liquid metal was then obtained by aspirating the liquid metal-Fe mixture with a syringe and rinsing it three times with deionized water.

## 3. Results

**3.1 Scheme of fabrication protocol**

3.1.1 Fundamental design and representative structure of LMCs

Inspired by the archetypal structure of natural cells, we built a biomimetic tri-layer architecture to show the essential design principle of LMCs (**Fig. 1B**). As a representative implementation, we selected EGaIn as the liquid metal core that provides dynamic responsiveness and adaptability characteristics. The surrounded solution is 0.5 mol/L NaOH solution, which acts as a cytoplasm, facilitating the exchange of ions and molecules with the external environment while maintaining the morphology and properties of the liquid metal. The outermost layer consists of a calcium lactate/sodium alginate-Poly (3,4-ethylenedioxythiophene): poly (styrene sulfonate) (CaL/SA-PEDOT:PSS) composite hydrogel membrane, chosen for its highly hydrophilic nature, tunable properties, facile crosslinking requirements, and excellent encapsulation capabilities.[32-34] This composite hydrogel combines the biocompatibility of CaL/SA hydrogel with high conductivity, optical transparency, flexibility, dispersibility, and environmental stability of PEDOT:PSS.[35-37] It offers enhanced mechanical strength and conductivity while maintaining essential membrane functions,[38] and its characteristic blue coloration facilitates structural observation.

3.1.2 Fabrication strategies and derivations

To achieve optimal integration of these layers with distinct viscosities into a unified structure while maintaining structural stability and functional flexibility, we developed a novel fabrication method, termed the layered cryogenic molding method (**Fig. 1C**). This scalable approach, versatile in both size and shape control, enables the fabrication of multi-layer structures with diverse morphologies at millimeter scales and above. The strategy allows for the rational selection and combination of different polymers and ionic solutions according to the specific properties of different metallic cores and functional requirements of LMCs, providing a robust platform for investigating the intrinsic properties and response behaviors of LMCs.

Utilizing this method, we first fabricate spherical LMCs as fundamental units and investigate their properties and functions. Upon rewarming, the high-density liquid metal core naturally settles at the bottom of the structure while maintaining its spherical geometry within a specific volume range (**Fig. 1D**). Based on the capillary height formula, one has:

$$a = \sqrt{\frac{\sigma_{\text{LM}}}{\rho_{\text{LM}} g}} \quad (1)$$

where $\sigma_{\text{LM}}$ and $\rho_{\text{LM}}$ denote the surface tension and density of liquid metal, g is the gravitational acceleration.[9]

We determine that stable spherical morphology is maintained when the volume of liquid metal ($V_{LM}$) is below 16.904 μL. Consequently, we standardize the liquid metal content from 0 to 20 μL for subsequent experiments. Furthermore, this fabrication approach enables us to achieve a diverse array of centimeter-scale structures with morphological characteristics reminiscent of various cellular architectures, including the amoeba, paramecium, triceratium, nucleated erythrocyte, epithelial cell, muscle cell, and neuron (**Fig. 1E**). The membrane thickness can be controlled by adjusting either the hydrogel solution concentration or the duration of the bilayer composite structure in the undercooled gel solution post-freezing. This thickness modulation directly influences the mass ratio between the membrane and the liquid metal core, thereby affecting the LMC's physical properties and kinetic parameters. This tunability enhances the versatility of the method and enables the customization of structural properties for specific applications.

### 3.2 Permeability of LMCs

The fundamental basis for liquid metal to establish a living matter system lies in its ability to respond to external stimuli. It requires membrane structures and intermediate layers within LMCs to enable efficient metallic core-environment interactions, making membrane permeability a critical parameter. To validate this transport property, we conducted experimental investigations by encapsulating concentrated pigment solutions within LMCs using the layered cryogenic molding method. It was observed that, over a short period, the pigment molecules gradually permeated the surrounding area, encapsulating the LMC and accumulating at the top due to density differential. (**Fig. 1F**). The underlying mechanism of this transport phenomenon lies in the porous structure and 3D interconnectivity of the pores of the hydrogel membrane, as evidenced by scanning electron microscopy (SEM) analysis (**Fig. 1G**). Quantitative analysis revealed a high porosity of 81.568% and an average pore diameter of 7.876 ± 1.910 μm. These structural characteristics provide the physical foundation for membrane permeability, as demonstrated by the outward diffusion of encapsulated pigment molecules. The demonstrated outward permeability establishes the membrane's capacity for material transport, with bidirectional exchange capabilities further validated through external chemical exposure experiments described in subsequent sections. This permeability confirms the LMC's potential for dynamic interaction with extra environments, setting the stage for exploring its responsive behaviors.

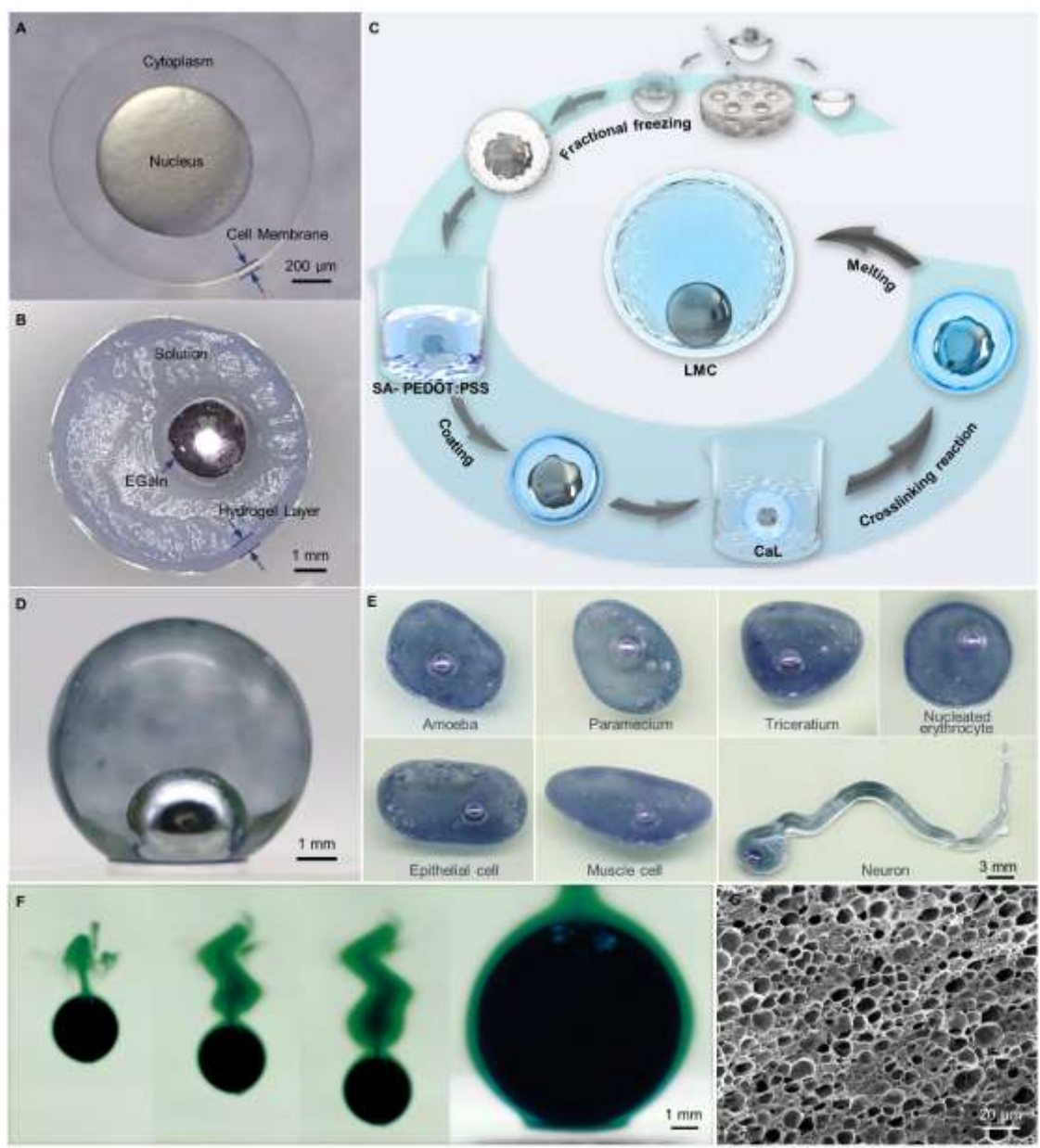

**Figure 1. Fabrication procedure scheme and typical structure demonstration of LMCs.** A. Zebrafish fertilized egg; B. Cross-section of a typical LMC in the frozen state; C. Schematic illustration of the fabrication protocol of LMCs; D. Side view of a typical LMC at room temperature;

E. A variety of LMCs with different morphologies manufactured by the layered cryogenic molding method; F. Diffusion of pigment solution encapsulated in the internal vesicle across the concentration gradient to the external solution environment; G. Microscopic structure of the cell membrane portion of the LMC cross-section.

## 4. Versatile Responses of LMCs under External Fields

Timely response to unknown changes and risks is fundamental for life systems to achieve survival, adaptation, and evolution. As a system parallel to the natural life system, the liquid metal living matter system must also confront complex and dynamic environments, rather than being perpetually protected within ideal laboratory conditions. As the foundation for establishing this artificial system, LMCs facilitate the transition of liquid metal from dependency on solution environments to functioning as independent entities in open-air conditions, thereby directly addressing challenges. Furthermore, the proportion of liquid metal serving as the core component represents a critical design parameter that fundamentally influences the properties and functions of LMCs. Therefore, we have systematically investigated the impact of liquid metal core content on LMC behavior (**Fig. 2A**). This section selects three representative external stimulus—temperature, pressure, and chemical fields—to explore how these cells adapt to natural environments (**Fig. 2B**). To effectively capture the dynamic responses of LMCs, we utilize electrical resistance as a key parameter, given that the constituent elements of the LMCs are all electrically conductive. It serves as a sensitive and quantifiable indicator of the structural and compositional changes within the cells when exposed to external stimuli. We discover that LMCs possess an inherent sensing capability.

### 4.1 Thermal response

Temperature fluctuations are inevitable in practical operational environments. To further investigate their behavior under varying thermal conditions, LMCs are maintained under controlled isothermal conditions across different temperature regimes from 15 ℃ to 40 ℃. By real-time monitoring of the average temperature at three points (top, middle, and bottom) and longitudinal resistance, we explore the impact of both internal liquid metal content and external temperature variations on the overall electrical properties of LMCs. Due to the high conductivity of liquid metal, the LMCs exhibit significantly lower resistance compared to the equal-volume hydrogel vesicles containing only solution. The resistance of LMCs exhibits an inverse relationship with temperature elevation (**Fig. 2C**). To analyze this phenomenon, we first examine the thermal effects on the conductivity of individual components. With increasing temperature, liquid metals show decreased electrical conductivity ($\sigma_{LM}$), resulting from enhanced electron scattering caused by intensified atomic thermal motion and wider interatomic spacing:

$$\sigma_{LM} = \sigma_0 - \alpha(T - T_m) + \beta(T - T_m)^2 \qquad (2)$$

where $\sigma_0$ is the electrical conductivity at the melting temperature, $T_m$ is the melting temperature, α and β are the temperature coefficient of variation.[40]

While for the solution, a decrease in its viscosity, an increase in the mobility of the ions, and an increase in the number of ions in the cytoplasmic solution caused by the increase in temperature will lead to an increase in the conductivity ($\sigma_S$):

$$\sigma_S = \sigma_0(1 + \delta(T - 25)) \tag{3}$$

where $\sigma_0$ is the conductivity of the solution in 25 °C, $\delta$ is the temperature coefficient of variation.[41]

Given that the maximum nuclear proportion in the entire cell is 10.233%, the alkaline solution, which performs cytoplasmic functions, continues to dominate the overall electrical properties, as its conductivity differs from that of EGaIn by approximately six orders of magnitude within the testing range.[42] As the nuclear proportion increases, the electrical conductivity of the composite structure progressively enhances under the same external conditions, accompanied by a gradual decrease in the rate of resistance change. This phenomenon indicates that the liquid metal is gradually regaining primary control over the LMC's overall properties, thereby enhancing the LMC's capacity to withstand thermal perturbations. LMCs' real-time sensing of temperature variations in both external and internal environments ensures stable and efficient operation of the liquid metal living matter system.

**4.2 Mechanical response**

Mechanical stress is ubiquitous in natural settings, whether from fluid dynamics, gravity, or physical contact with surrounding materials. To investigate their mechanical resilience, LMCs are subjected to uniform vertical compression at a rate of 2.4 mm/min. During this time, they rapidly detected the increasing pressure and exhibited enhanced electrical conductivity properties as a characteristic response **(Fig. 2D₁)**. Within the compression range considered in this study, the resistance decreases with increasing liquid metal content, showing a gradually diminishing rate of change. To elucidate the underlying mechanism of these behaviors, we developed a simplified geometric model to analyze the electrical resistance change during the compressive deformation of cells **(Fig. 2D₂)**. Following the principle of charge transport along electric field lines, the overall resistance can be simplified into four series-connected components based on the current path. The first component is a spherical liquid metal core with radius $R_{LM}$. The second component is a truncated cone-shaped solution section, having a total height of $H$ together with the first part. Its upper surface, with radius $R_T$, represents the circular contact area between the solution and the hydrogel membrane interfacing with the top electrode and its side extensions tangent to the core. The remaining two components represent the resistance of the hydrogel in contact with the top and bottom electrodes, with contact areas $A_T$ and $A_B$ and thickness $t_h$. According to this model, the resistance of the LMC ($R_{LMC}$) under compression can be expressed as:

$$R_{LMC} = \frac{1}{4\pi\sigma_{LM}R_{LM}} + \frac{1}{\sigma_s \pi R_T \frac{R_T}{H - 2R_{LM}} + \tan(\sin^{-1}\frac{R_{LM}}{H - R_{LM}})} + \frac{t_h}{\sigma_h}\left(\frac{1}{A_T} + \frac{1}{A_B}\right) \tag{4}$$

where $\sigma_s$, $\sigma_{LM}$, and $\sigma_h$ is the electric conductivity of the solution, liquid metal, and hydrogel, respectively.

As the compression distance increases, this model predicts an overall decrease in electrical resistance. It is primarily attributed to the reduction in cell height $H$, the expansion of the $R_T$, $A_T$, and $A_B$. The theoretical predictions demonstrate excellent agreement with our experimental observations. Additionally, the resistance and its change rate gradually decrease with increasing nucleus proportion under equivalent compression distances, which can be explained by the monotonic variation of $R_{LMC}$ and its derivative with increasing $R_{LM}$ within the investigated range (**Eq. 4**). This further demonstrates that the incorporation of liquid metal enhanced the environmental adaptability of the overall structure. No structural failure occurs during the compression tests, showing the structural integrity and deformation ability of these composite structures. These mechanical properties enable LMCs to extend their adaptability, the physical and functional reconfigurability, across diverse pressure environments under external fields.

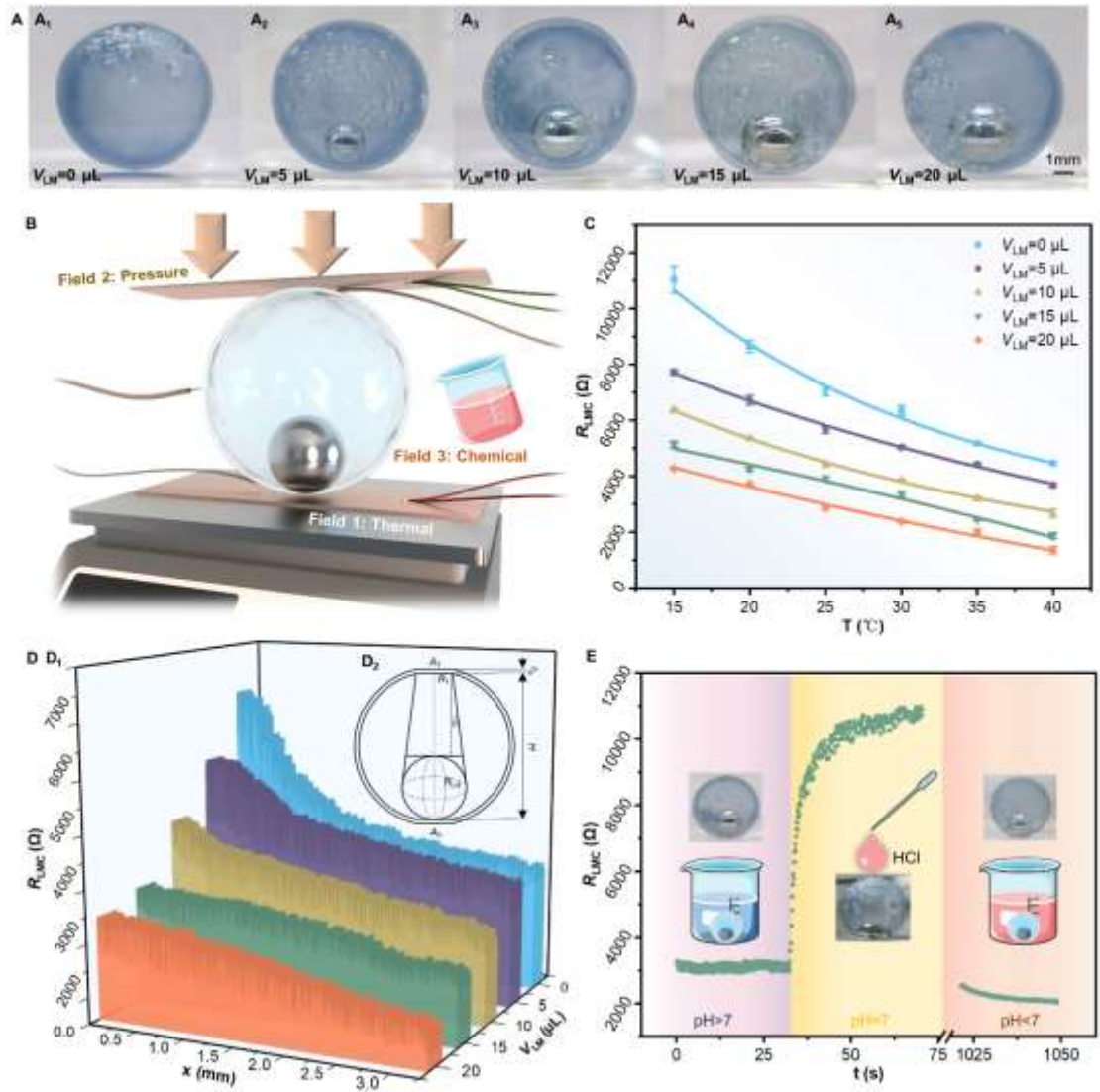

**Figure 2. Response under pressure, temperature, and chemical fields of LMCs.** A. LMCs with varying nuclear proportions ($V_{LM}$ =0, 5, 10, 15, and 20 μL) under ambient temperature and atmospheric conditions; B. The LMC placed in a simulated complex environment; C. Resistance changes of LMCs in the temperature range from 15 to 40 ℃; D. Variations in the electrical properties of LMCs during longitudinal compression from 0 to 3.27 mm at 26 ℃; E. Instantaneous effect of external pH changes on the longitudinal resistance of a LMC ($V_{LM}$=15 μL) at 26 ℃.

### 4.3 Chemical response

In addition to physical field variations, chemical gradients in the natural environment are inevitable factors to which living matters must adapt. Therefore, we further investigate whether the LMCs could adapt to rapid changes in external chemical environments by maintaining their dynamic equilibrium. When exposed to the hydrochloric acid solution (HCl, 0.5 mol/L, 350 μL) with equivalent concentration and volume to the internal cytosol solution, the LMC ($V_{LM}$ = 15 μL) exhibits distinct morphological and physicochemical responses (**Fig. 2E**). Due to the neutralization reaction, ion concentration within the cytoplasmic region reduces,[43] and an oxide layer develops on the surface of the liquid metal core.[44] Thus, the overall electrical resistance of the LMC rapidly increases. Benefiting from the excellent permeability of the membrane structure, the ionic concentration within the intramembrane area gradually equilibrates with the external HCl solution after 20 minutes. HCl solution effectively removes the oxide layer from the liquid metal surface, similar to the effect of NaOH solution, restoring its spherical morphology and electrical properties.[45] Since the conductivity of 0.5mol/L HCl solution is slightly higher than that of NaOH solution (17.8 S/m for HCl and 9.31 S/m for NaOH),[42] the overall resistance of the LMC within excess hydrochloric acid solution decreases significantly compared to that in the neutralization phase, and is slightly lower than the initial resistance in NaOH solution. These response patterns directly reflect the combined influence of the conductivity of the solution and liquid metal on the overall resistance of the LMC, as predicted by the theoretical resistance model (**Eq. 4**).

### 5. Motions of LMCs

Beyond static responses, directional movement capabilities are essential for artificial living systems to interact with and adapt to their environment and perform complex tasks. For liquid metal living matter, the development of movement strategies should capitalize on their unique physicochemical properties, particularly their exceptional responsiveness to electromagnetic fields.[46,47] The capacity of LMCs to approach beneficial conditions and avoid adverse environments is essential for their environmental adaptation and functional performance, which can be achieved through controlled responses to external fields. This section focuses on the changes in the position and morphology of LMCs under the tuning of electric and magnetic fields in a solution environment.

**5.1 Electrical tuning behavior**

Electric fields permeate natural environments, influencing various cellular behaviors and biological processes.[17] The electrical response behavior of LMCs represents another sophisticated manifestation of their field-responsive properties. We explore the electric tuning behavior of LMCs within a 17 mm wide and 50 mm length quartz channel containing solution at a depth of 13 mm (**Fig. 3A₁**). It is observed that LMCs exhibited significant directional movement in the applied voltage of higher than 7.91 V ($V_{LM}$=20 μL). After an initial acceleration period, LMCs maintained uniform velocities while moving along the electric field lines from cathode to anode.

To further investigate the relationship between LMCs' composition and electrical tuning behavior, we varied the $V_{LM}$ from 5 to 25 μL. Our results reveal that both the activation voltage and movement speed were significantly influenced by $V_{LM}$. Specifically, the minimum voltage required for initiating movement shows a negative linear correlation with $V_{LM}$ within a certain range (**Fig. 3B**). Under a fixed potential difference of 12 V, the steady-state velocity of LMCs shows pronounced growth in the 5 to 15 μL range of $V_{LM}$. However, the velocity increase becomes slower beyond 15 μL, likely due to the combined effects of increased mass and droplet deformation.

To elucidate the underlying mechanism of this field-induced directional motion, we develop a theoretical framework focusing on the driving force analysis of LMCs. Cells containing only solution without LM cores showed no movement under the electric field, demonstrating that the motion of the LMC is primarily driven by the nucleus. Therefore, the force analysis focuses mainly on the LMC nucleus (**Fig. 3A₂**). It is primarily subjected to forces resulting from viscous drag caused by the surrounding solution ($f_\eta$), and frictional forces from the hydrogel membrane at the bottom ($f_\beta$), and surface tension imbalance arising from the electric double layer presence between the two hemispheric sides of a LMC parallel to the direction of the electric field ($F_\gamma$).[48-50]

$F_\gamma$ is the main driving force in elucidating the intrinsic mechanisms governing the LMCs' dynamic behavior. According to our force analysis model of the LMC, this driving force can be expressed as:

$$F_\gamma = \frac{4 q_0 \cdot U_E}{L_E \cdot \left( \frac{1}{\pi r_{LM}^2} - \frac{2 h_1 h_2}{3} \right)} \quad (5)$$

where $q_0$ is the amount of charge stored in the electrical double layer at the liquid metal surface when treated as a capacitor, $U_E$ is the potential difference between the electrodes on either side, $L_E$ is the distance between two electrodes, $r_{LM}$ is the radius of the liquid metal droplet, $h_1$ and $h_2$ are the width and height of the channel filled with liquid, respectively.

Mathematical analysis of the equation reveals that the driving force exhibits positive correlations with $U_E$ both and $r_{LM}$, while showing an inverse relationship with $L_E$. These theoretical predictions align well with our experimental findings, where larger droplets demonstrated enhanced mobility due to increased driving forces, while smaller droplets required higher activation voltages for effective movement. The theoretical model not only provides a quantitative basis for understanding the size-dependent electrical response behavior but also provides insights into the key parameters governing LMCs' motion.

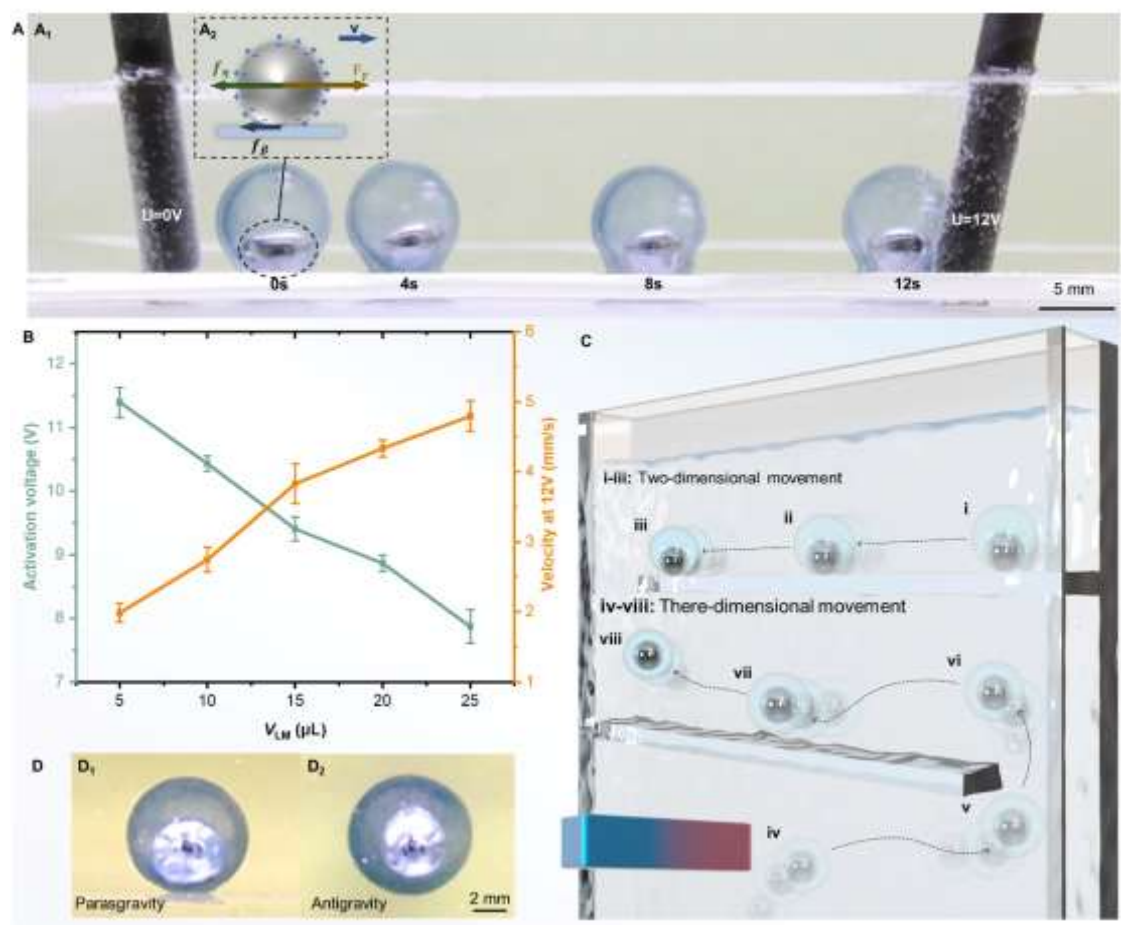

**Figure 3. Directional motion characteristics of LMCs.** A. The movement process of a LMC ($V_{LM}$=20 μL) over a period of 12 seconds at a potential difference of 12 V between two electrodes spaced 5 cm apart; B. Effect of volume of EGaIn in LMCs on activation voltage and average motion speed; C. Two-dimensional and three-dimensional motion of a magnetic LMC guided by a magnet in a vertical plane; D. Two different motion morphologies of a magnetic LMC: $D_1$. Dropping and $D_2$. Suspension.

### 5.2 Magnetic tuning behavior

Magnetic fields, as one of the fundamental physical forces on Earth, play crucial roles in biological navigation and spatial orientation.[51,52] The diversity of the liquid metal composites further extends the functional expandability of LMCs. To expand the locomotion capabilities of LMCs beyond electric field response, we incorporate magnetic responsiveness by replacing EGaIn with liquid metal ferrofluid through the layered cryogenic molding method. The magnetic field-driven locomotion is investigated using a custom-designed experimental setup comprising two platforms at different heights, with the magnetic LMC immersed in a 0.5 mol/L NaOH solution (**Fig.**

**3C**). The magnetic LMCs exhibit distinct locomotion characteristics with remarkable multi-dimensional mobility, including lateral movement, anti-gravity motion, and corner-crossing motion guided by a NdFeB magnet. The sequential motion trajectory (positions i-iii) illustrates one-dimensional lateral movement confined to a horizontal platform. The liquid metal core preserves high surface tension and maintains at the bottom because of gravitational effects (**Fig. 3D$_1$**). When the magnetic force in the vertical direction counterbalances the gravitational force, the LMC achieves controlled suspension. Sequential motion trajectory (positions iv-viii) shows the flexible two-dimensional motion of the magnetic LMC under the guidance of the magnet. It enters a suspended state, with the liquid metal core stabilized centrally within the cell's inner membrane to the vertical wall (**Fig. 3D$_2$**). This multidimensional manipulation via magnetic fields demonstrates that LMCs possess significant potential for functional evolution through strategic materials integration. Such capability facilitates the development of responsive systems capable of controlled directional movement under the tuning of various external fields including light, sound, thermal fields, etc.

## 6. Capability of Resisting Low-Temperature Conditions

While the previous section examined LMC behavior under moderate temperature variations (15-40°C), extreme thermal conditions present more significant challenges for functional materials. Understanding the stability and performance retention of LMCs under extreme low-temperature conditions is crucial for evaluating their operational reliability across extended environmental ranges. Here, we explore their response to rapid cooling from above 0°C to -13°C, focusing on the property changes of both integral and individual components. Due to the high water content (95.39%) and the permeability of the hydrogel membrane, it is considered as a unified system with the solution, and both exhibit synchronized phase transitions.

To evaluate the effect of liquid metal on cold resistance, this section compares LMCs with pure vesicles of the same size containing only solution as a control. The control group exhibits distinct thermal behavior characterized by a temperature plateau at -5°C, with resistance changes showing a negative correlation with temperature variations before and after this point (**Fig. 4A**). In stage i, the hydrogel maintains its gel state while the solution remains liquid. A significant 103.143% step increase in resistance marked the phase transition between two distinct stages. Then, both components solidify in stage ii.

In contrast, LMCs demonstrate a more complex cooling process (**Fig. 4B**). During stage iii, resistance increases with decreasing temperature while maintaining the cell membrane's gel state and liquid phases of both solution and liquid metal. The cooling rates remain similar through stages i and iii due to the low volume fraction of liquid metal within the LMCs (10.234%) and extensive cooling source. Similarly, there is no significant influence on the phase transition time of the outer components. A 98.439% step increase in resistance marks the transition to stage iv, where all components except liquid metal solidified.

Stage v is characterized by liquid metal reaching its phase transition point at -7.3 °C, releasing latent heat that elevates the surrounding temperature. It temporarily reverts nearby components to their stage iii. Finally, in stage vi, complete solidification occurs with stabilized resistance. Due to the incorporation of liquid metal with high specific heat capacity, the release of supercooling and phase change latent heat elevates the overall temperature descent curve, thereby delaying the occurrence time of the minimum temperature.[53]

Overall, the incorporation of the metallic core fundamentally modifies the system's thermal behavior through phase change-induced thermal buffering. During the initial cooling process, the latent heat release rapidly elevates the temperature to 90.76% of that observed in the blank control group and minimizes to 73.07%. This thermal modulation capability enhances the robustness of LMCs, providing improved structural integrity and functional preservation under extreme environmental conditions, which is the critical feature for practical applications such as cryopreservation or deployment in harsh environments.

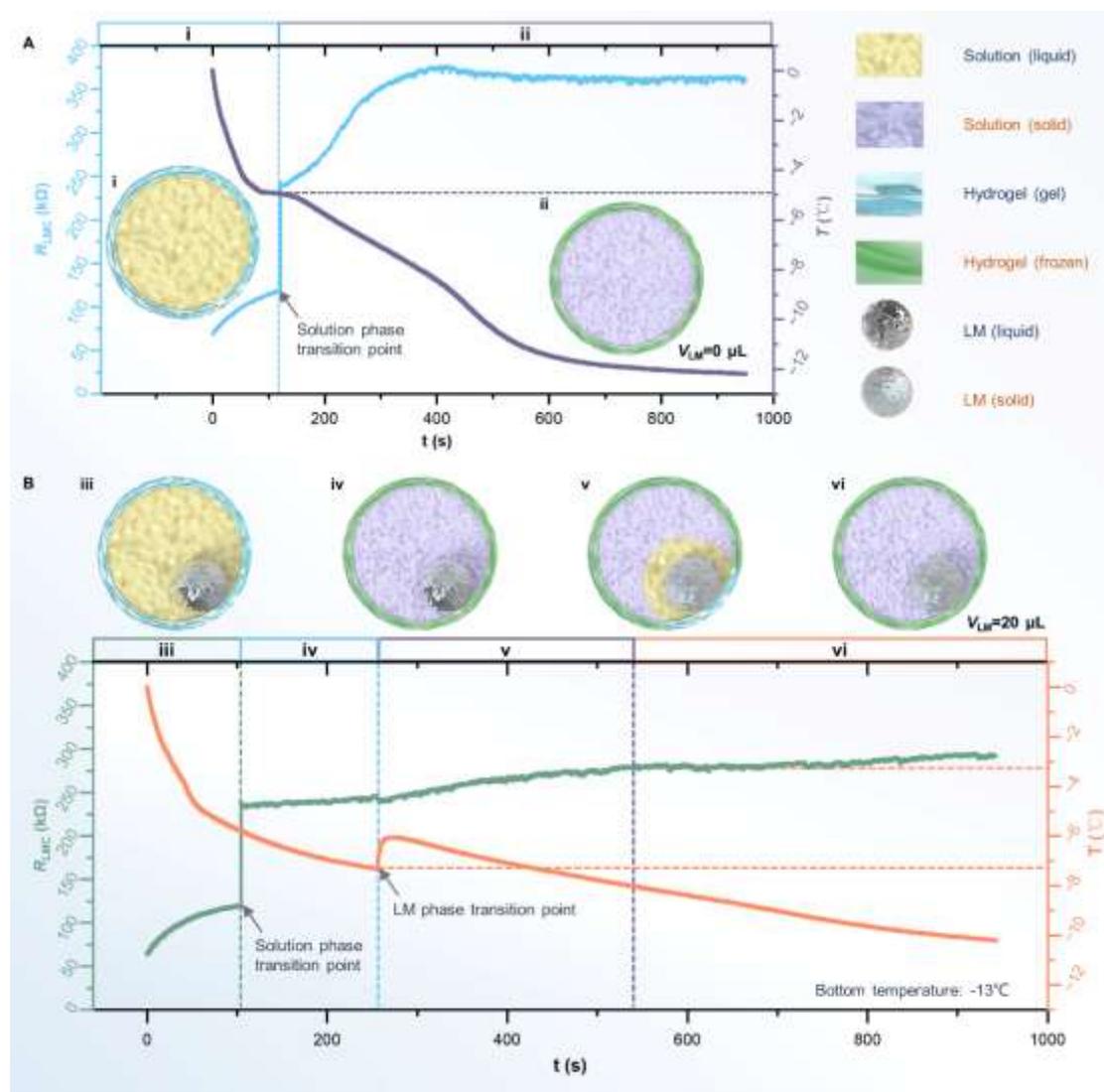

**Figure 4. Changes in the response characteristics of LMCs under extreme low-temperature**

**conditions.** A-B. Resistance and temperature change curves, as well as phase transition schematic during the cooling process of: A. pure vesicle without the addition of liquid metal ($V_{LM}$=0 μL); B. LMC ($V_{LM}$=20 μL).

## 7. Dynamic Control of LMCs in Deformed Channels

The emergence of LMCs represents a pivotal step in transforming liquid metal-based living matter systems from conceptual frameworks into practical realities. They provide an extensible, multifunctional platform that potentially expands the structural and functional possibilities of liquid metals across various domains. To demonstrate this potential, we showcase their capabilities in deformed channels (**Fig. 5A**), where they function as sensors, valves, and delivery systems, integrating electrical conductivity, magnetic responsiveness, and morphological adaptability. This demonstration highlights their environmental adaptability and multifunctionality, illustrating how LMCs can operate effectively in dynamic environments.

Here we designed a converging-bifurcating channel system with a width gradient from 12 mm to 6 mm to validate the potential of magnetized LMCs in flow control applications. **Fig. 5B** shows the experimental phenomena and corresponding deformation of LMCs and flow states of this system. In the initial phase, a magnetic LMC ($V_{LM}$=20 μL) is positioned at the channel's widest cross-section, and an aqueous solution is introduced at a flow rate of 3.5 mL/s (**Fig. 5B$_1$**). The LMC exhibits controlled migration along the flow direction with position correlation exceeding 0.99 relative to magnetic guidance, demonstrating remarkable deformability as it adapts to the narrowing channel geometry while maintaining forward momentum. Upon reaching a channel segment matching its volume or the bifurcation, the LMC establishes complete occlusion, effectively halting the fluid flow (**Fig. 5B$_2$**). Due to the fixed volume constraint of the system and the unidirectional direction of flow, this blockage creates a significant hydraulic pressure differential across the LMC, generating a 5 mm liquid level difference within 7 seconds, evidenced by distinct liquid levels on either side.

The system's reversibility is demonstrated through magnetic guidance, where the LMC can be manipulated to restore flow patency. Under magnetic control, the magnetized LMC undergoes controlled movement (**Fig. 5B$_3$**), allowing fluid redistribution through the newly created channel space. The liquid level difference between the opposing sides of the LMC demonstrates a decrease over time. The magnetic force exerted on the ferrofluid core effectively overcomes the hydrodynamic forces, enabling precise positional control of the LMC. The system ultimately restores normal flow patterns (**Fig. 5B$_4$**). The reversibility and structural robustness of this approach were validated through 20 consecutive blockage-release cycles performed within 20 minutes, with the LMC maintaining functional integrity throughout the entire operation.

Additionally, the photothermal effect and photoacoustic effect of liquid metal enable advanced multi-modal wireless sensing and remote control strategies in complex flow channels while providing imaging capabilities. [54,55] The LMCs, achieving dynamic control through these liquid metal cores, are integrated with two additional material delivery modalities: hydrogel matrices and solution.[56-58] Leveraging the extensible material properties of the layered cryogenic molding method, channel recanalization and targeted material delivery can be achieved through the controlled degradability of hydrogels. [59-61] LMCs create a versatile platform for complex flow channel management applications. Looking forward, applications in increasingly complex scenarios will necessitate the concurrent optimization of material composition, structural design, and process refinement to achieve enhanced functionality and scalability.

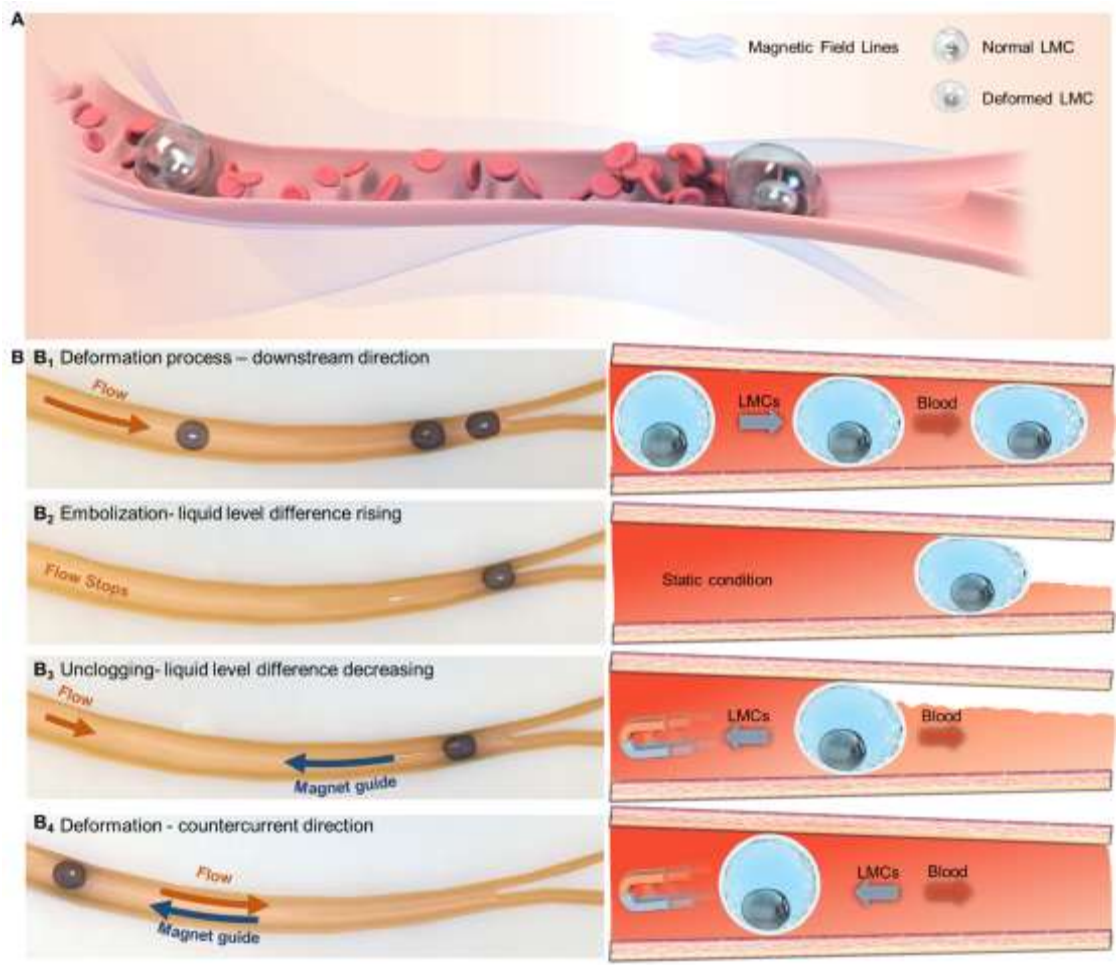

**Figure 5. Movement and deformation of LMCs in a deformed channel.** A. LMCs demonstrate promising potential for flow control applications. B. The magnetic LMC performs self-deformation and blockage functions within the channel and achieves reverse flow motion under the influence of an external magnetic field: $B_1$. The LMC moves along the direction of the fluid flow and deforms to adapt to the surrounding environment; $B_2$. Upon reaching the maximum deformation, the LMC becomes stationary and fills the flow channel, resulting in a blockage; $B_3$. Under the guidance of a magnet, the LMC moves in the direction opposite to the flow, accompanied by shape recovery; $B_4$. States of the LMC and flow are restored.

# 8. Conclusion

In this study, we systematically present the concept, preliminary theory, and technology of LMCs as the fundamental structural and functional units of liquid metal living matter and successfully establish their fundamental architecture using a systematic bottom-up fabrication strategy. Through the layered cryogenic molding method, we develop a multi-layer and multi-phase cell mimic wherein the liquid metal core is the nucleus for responsive functionality. The reducing electrolyte solution works as the cytoplasm, simultaneously maintaining the metallic core's reductive state advantages while facilitating information and material exchange between the internal and external compartments. The outer membrane is an alginate-based composite hydrogel with a porous architecture, providing structural support and permeability for molecular and ionic transport. This LMC unit with a tri-layer architecture effectively addresses previous limitations of liquid metal applications in scale and environment, maintaining the morphological and physicochemical stability of liquid metals.

As the basic unit of liquid metal living matter, the response properties of such LMCs to environmental stimuli are first investigated. Through systematic measurement and analysis of characteristic resistance, we find that LMCs possess remarkable sensing capabilities and environmental adaptability. They exhibit instantaneous and well-regulated resistance changes in response to varying temperature, pressure, and chemical fields. To clarify the response mechanisms induced by different environmental stimuli, we establish a geometric resistance model that links electrical resistance to the structural deformation of the LMC. Both experiment results and the resistance model demonstrate that the integration of the liquid metal core into the LMC structure significantly enhances its perceptual ability and intrinsic homeostasis, with the model successfully capturing the key scaling behaviors observed across different compression states.

The multi-field tuning capability of LMCs builds the foundation for developing various external field responsive behaviors, the most typical of which are electrical and magnetic tuning behavior. Therefore, the movement controllability of LMCs under electric and magnetic fields is studied with varying liquid metal nuclei proportions. It reveals that LMCs are capable of controlled directional movement under electric fields, and LMCs with higher liquid metal adding ratios show more sensitive tuning ability, manifested by higher moving speed and lower activation voltage. By modeling the force analysis of LMCs, we found that the key to the LMC motion regulation by electric field lies in the pressure difference caused by the surface tension gradient generated across the liquid metal nucleus. The magnitude of this force is closely related to the structural ratio between the solution and the liquid metal nucleus within the hydrogel membrane. Mechanical modeling demonstrates that in the scale range of liquid metal cells that we have explored, the increase in the proportion of liquid metal nuclei has a significant gain effect on enhancing the driving force generated by the cell subjected to the electric field. In addition, comparative experiments with blank groups confirm that pure solution vesicles without added liquid metal nuclei cannot be driven by electric fields.

Furthermore, by virtue of the extensibility of liquid metal composite material systems, the LMC can be functionalized with magnetic-tuning ability by incorporating ferromagnetic particles into the liquid metal core. The magnetic LMCs demonstrate precise spatiotemporal control over their locomotion and positioning in three-dimensional space. Such a contactless motion control approach, coupled with the fully flexible and biocompatible composite structure of LMCs, as well as the viability and versatility of the structural design methodology, jointly endows the LMCs with navigational ability through adaptive shape modulation. These unique structural and responsive characteristics of LMCs make them particularly promising for applications in non-uniform channels where wired control is not feasible. Finally, as a practical illustration, we design a converging-bifurcating channel system with a width gradient to validate the potential of magnetized LMCs as sensors, valves, and delivery systems.

In summary, the conceptual establishment, response characterization, and application demonstration of LMCs in this paper provide an initial insight into the unique features of the liquid metal living matter system. By establishing the foundational principles of multi-field responsiveness and functional integration in these engineered constructs, the finding offers a conceptual framework for future explorations in adaptive materials. However, numerous aspects remain to be explored to further enhance the intelligence level and expand the survival scenarios. It is crucial to optimize LMCs' long-term stability, explore diverse material combinations, and develop more sophisticated multi-field control strategies. Furthermore, investigating collective behaviors and interactions among identical LMCs would facilitate their assembly into higher-order liquid metal functional structures. Through the controlled fabrication processes we established, developing diverse LMCs with distinct shapes, functions, and adaptabilities would advance this system toward ever higher level complexity. As reliable basic building blocks, the LMCs offer a solid foundation for the assembly of more sophisticated liquid metal living matter, potentially leading to unprecedented applications in various fields.

## Declaration of interests

The authors declare no competing interests.

## Acknowledgements

This work was partially supported by the National Natural Science Foundation of China under Grants No. 12402324 and No. 91748206.

## Author contributions

Conceptualization: JY.L. and J.L.; Methodology: JY.L., X.Z., and J.L.; Investigation: JY.L., R.X., J.W., and Y.S.; Data curation: JY.L. and M.Q.; Formal analysis: JY.L., Q.M., and X.Z.; Validation: JY.L., Z.X., and Y.B.; Visualization: JY.L., M.G., X.Z., and J.L.; Writing—original draft: JY.L.; Writing—review and editing: X.Z., and J.L.; Supervision and funding acquisition: X.Z. and J.L.; All authors read and approved this manuscript.